\documentclass{article}

\title{Extension of Sinkhorn Method: Optimal Movement Estimation of Agents Moving at Constant Velocity}

\usepackage{graphicx}

\usepackage{lipsum}
\usepackage{amsmath}
\usepackage{amsfonts}
\usepackage{float}
\usepackage{lineno,hyperref}
\usepackage{caption}
\usepackage{mathptmx}       
\usepackage{helvet}         
\usepackage{courier}        
\usepackage{type1cm}        
\usepackage{makeidx}         
\usepackage{multicol}        
\usepackage[bottom]{footmisc}
\usepackage{amsmath}
\usepackage{amssymb}
\usepackage{amsfonts}
\usepackage{usebib}
\usepackage{authblk}
\usepackage{blindtext}

\author[1]{Daigo Okada \thanks{dokada@genome.med.kyoto-u.ac.jp}}
\author[1]{Naotoshi Nakamura \thanks{nnakamura@genome.med.kyoto-u.ac.jp}}
\author[2]{Takuya Wada \thanks{peacefield.taku3@gmail.com}}
\author[2]{Ayako Iwasaki \thanks{iwasaki.ayako.38n@st.kyoto-u.ac.jp}}
\author[1]{Ryo Yamada \thanks{Corresponding Author: ryamada@genome.med.kyoto-u.ac.jp}}
\affil[1]{Graduate School of Medicine, Kyoto University,  Syogoin Kawaharachoh 53, Sakyo, Kyoto, Kyoto Prefecture, Japan}
\affil[2]{Department of Medicine, Kyoto University,  Syogoin Kawaharachoh 53, Sakyo, Kyoto, Kyoto Prefecture, Japan}

\begin{document}

\maketitle

\begin{abstract}
In the field of bioimaging, an important part of analyzing the motion of objects is tracking. 
We propose a method that applies the Sinkhorn distance for solving the optimal transport problem to track objects.
The advantage of this method is that it can flexibly incorporate various assumptions in tracking as a cost matrix.
First, we extend the Sinkhorn distance from two dimensions to three dimensions.
Using this three-dimensional distance, we compare the performance of two types of tracking technique, namely tracking that associates objects that are close to each other, which conventionally uses the nearest-neighbor method, and tracking that assumes that the object is moving at constant velocity, using three types of simulation data. 
The results suggest that when tracking objects moving at constant velocity, our method is superior to conventional nearest-neighbor tracking as long as the added noise is not excessively large. 
We show that the Sinkhorn method can be applied effectively to object tracking. Our simulation data analysis suggests that when objects are moving at constant velocity, our method, which sets acceleration as a cost, outperforms the traditional nearest-neighbor method in terms of tracking objects. To apply the proposed method to real bioimaging data, it is necessary to set an appropriate cost indicator based on the movement features.

\end{abstract}

\section{Introduction}

In the field of bioimaging, an important part of analyzing the motion of objects is tracking \cite{biotracking1,biotracking2,biotracking3}. The tracking process can be described as follows. Images taken at fixed time intervals contain many objects. The goal is to identify which signals correspond to which object at the next time point. This task is important for bioimaging analysis, such as the analysis of microscopy videos, because it is indispensable for analyzing the motion of objects from image data taken at fixed time intervals. However, automatic tracking is difficult. Many types of algorithm have been proposed for this task \cite{tracking1,tracking2}, including the nearest-neighbor method \cite{NN}, probabilistic data association\\
\cite{PDA}, and multiple hypothesis tracking \cite{MHT}.

Nearest-neighbor algorithms, the most simple of tracking methods, are used for live-cell tracking in the field of bioimaging analysis \cite{nnforbio}. These algorithms associate objects that are close to each other. Although this is a simple task, the performance of nearest-neighbor algorithms is inadequate when the objects are crowded together or their movement distance is long. These difficult conditions are common in bioimaging data. In this study, we extend the nearest-neighbor method. Because nearest-neighbor tracking can be considered as an optimal transport algorithm, we adopt the Sinkhorn method \cite{sinkhorn}, an optimal transport algorithm, to modify nearest-neighbor tracking.

In this research, we apply the Sinkhorn method to object tracking. This allows us to perform tracking using various transport costs based on a model of object behavior. For tracking, we do not have to associate the nearest objects at two consecutive time points; we can associate objects so that their trajectories are smooth. A smooth trajectory means that changes in velocity are small, or that the objects are moving at constant velocity. Therefore, we use three consecutive time-point images to measure changes in velocity, weigh the changes as a cost, and optimize the combination of spots in the three time-point images using Sinkhorn regularization. In the following sections, we describe the notation of the nearest-neighbor-based Sinkhorn approach and the extension of the Sinkhorn method to optimization for objects moving at constant velocity, followed by methods for generating simulation datasets. We then compare the performance of two Sinkhorn-based methods, namely the nearest-neighbor method and the proposed method, using the datasets.

\section{Theory/calculation}
\label{sec:2}
\subsection*{1. Overview of optimal transport problem and Sinkhorn method}
Optimal transport has been investigated as a major problem in information science. It can be applied to various fields \cite{otbase}, including tracking in imaging analysis \cite{ot}. The optimal transport problem is as follows. Items are distributed in spots, which we call sources. We want to move the items to new spots, which we call targets. What we want to know is how many items should be moved from where to where. We want to move the items with minimum cost. The cost is the sum of the products of the volume of items and the distance between spots. The optimal transport problem can be defined mathematically as follows. The source and target are discrete mass vectors $r$ and $c$, which satisfy the definition of a discrete probability distribution. Transportation, how many items are moved from where to where, is expressed as matrix $P$, whose elements represent the amount of movement from each source to each target. The elements are non-negative and their row and column sums are the source and target vectors, respectively. With the number of objects denoted as $d$, $P$ is a $d \times d$ square matrix, where the numbers of rows and columns of $P$ correspond to the numbers of elements of $r$ and $c$, respectively.
$P$ is defined by the following equation:
\begin{equation}
U\left(r,c\right) := \{ P \in \mathbb{R}_+^{d \times d} | P \vec{1}_d = r, P^T \vec{1}_d = c \}
\end{equation}
where $\mathbb{R}_+^{d \times d}$ represents a set of matrices, whose elements are all non-negative, and $\vec{1}_d$ represents a vector, whose elements are all 1. Each element of $P$ represents the transportation of an object from one spot in $r$ to one spot in $c$. Therefore, using cost matrix $M$, whose size is the same as that of $P$ and whose elements represent the unit cost of the corresponding transportation, $\langle P,M \rangle = \sum_{i,j} p_{ij} m_{ij}$ is the Frobenius inner product of $P$ and $M$ where $p_{ij}$ and $m_{ij}$ are (i,  j)th element of $P$ and $M$, respectively, which represents the total cost. The optimal transportation matrix $P^{*}$ is defined as follows:
\begin{equation}
\newcommand{\argmin}{\mathop{\rm arg~min}\limits}
d_{M}\left(r,c\right)  = \langle P^{*}, M \rangle, \ \ \ \  \mathrm{where} \ \ \  P^{*} = \argmin_{P \in U\left(r,c\right)} \langle P, M \rangle .
\end{equation}
The transport distance $d_{M}\left(r,c\right)$ \cite{transd} calculated using a cost matrix whose elements are the distances of corresponding transportation of spots can be used as the distance between $r$ and $c$. Sinkhorn proposed an algorithm for the optimization of this distance \cite{sinkhorn}. Because the search space of this optimization, $U\left(r,c\right)$, is large, Sinkhorn added a regularization term based on entropy and transformed the minimization of the transport distance into the minimization of the Sinkhorn distance $d_M^{\lambda}\left(r,c\right)$, as defined below. This regularization makes the area smaller and thus easier to search. The Sinkhorn distance is defined as:
\begin{equation}
\newcommand{\argmin}{\mathop{\rm arg~min}\limits}
d_{M}^{\lambda}\left(r,c\right)  = \langle P^{\lambda}, M \rangle, \ \ \ \ \mathrm{where} \ \ \ P^{\lambda} = \argmin_{P \in U\left(r,c\right)} ( \langle P, M \rangle -\frac{1}{ \lambda} h\left(P\right) ).
\label{eq:01}
\end{equation}
where $h\left(P\right)$ is the entropic term $\sum_{i,j=1}^d p_{ij}\log p_{ij}$ and $\lambda$ is a user-defined regularization parameter. A larger value of $\lambda$ leads to a wider search range. If $\lambda$ is set to infinity, the transport distance and the Sinkhorn distance are equal. Using this method, the nearest-neighbor tracking of multiple objects can be regarded as an optimal transport problem where the distance between objects in the image and objects at the next time point is the transport cost.

\subsection*{2. Tracking Using Sinkhorn Method}
\label{subsec:2}
Here, we describe nearest-neighbor tracking using the Sinkhorn method. The input data are the position coordinates of particles at two time points ($t$ and $t+1$). To apply optimal transport to the tracking problem, we prepare two vectors, A and B, whose elements are all of the form $\frac{1}{n}$, where $n$ is the number of objects. A and B correspond to the source and target vectors, respectively, of the optimal transport problem. The cost is the distance between each pair of objects at time points $t$ and $t+1$. Cost matrix $M$ is the distance matrix between the objects at time point $t$ and those at time point $t+1$. If the time is equally spaced, this distance can be thought of as a speed. Transportation between two points is estimated using the Sinkhorn method. Transportation matrix $P$ and cost matrix $M$ are both two-dimensional (2D). The optimization is described by Eq. (1) in subsection 1. We call this situation "speed cost". In this case, optimal transport matrix $P$ represents nearest-neighbor matching.

We assume that the objects are moving at constant velocity. Then, we calculate the acceleration from the data obtained at three time points and use it as the cost. Table 1 shows our extension of the Sinkhorn distance from two dimensions to three dimensions.

%
\begin{table*}
\caption{Comparison of original 2D Sinkhorn distance and proposed 3D extension}
\label{tab:1}       
%
%
\begin{tabular}{p{4.5cm}p{5cm}p{4cm}}
\hline
 & Original Sinkhorn distance & Proposed extension  \\
\hline \hline
Considered time points&$t$, $t+1$ & $t$, $t+1$, $t+2$  \\
Estimated movement& $A_i \rightarrow B_j$ & $A_i \rightarrow B_j \rightarrow C_k $ \\
Dimension of $P$ and $M$ & 2 & 3  \\
Element of $M$ as cost index &  $||b_j - a_i||$ &  $||\left(c_k - b_j\right) - \left(b_j - a_i\right)||$  \\
 \hline
\end{tabular}
\end{table*}

In summary, the proposed extension uses three time points ($t$, $t+1$, and $t+2$) rather than two. Moreover, our estimation is not for pairs but for triples. The transportation and cost of 2D matrices need to be extended to three-dimensional (3D) arrays. However, the optimization formula is identical. The cost is acceleration rather than speed, or the second difference rather than the first difference. Our method returns the optimal triples rather than pairs, and the output is 3D optimal transport array $P$. Therefore, we compress the information in the triples down to pairs with the following formula and adopt the resulting matrix as the optimal transport matrix. 
\begin{equation}
p'_{ij} = \sum_{k=1}^n p_{ijk}
\end{equation}
where $p'_{ij}$ is (i, j)th element of the compressed 2D cost matrix and $p_{ijk}$ is (i, j, k)th element of original 3D cost array $P$.
Figure 1 shows the difference between speed cost tracking and acceleration cost tracking.

\begin{figure}[tbp]
\includegraphics[scale=.7]{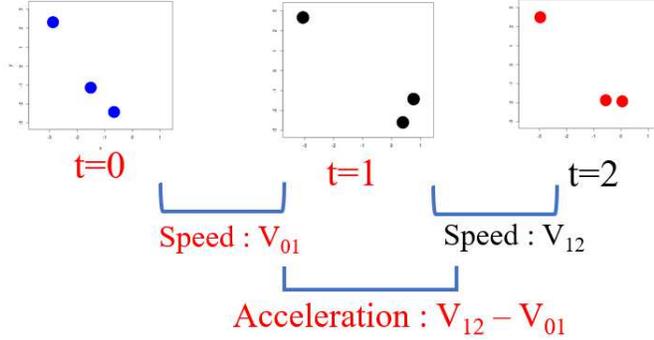}
%
%
\captionsetup{labelformat=empty,labelsep=none}
\caption{Fig. 1: Comparison of speed cost tracking and acceleration cost tracking. For speed cost tracking of objects at $t$ and $t+1$, Speed $V_{01}$ is used. For acceleration cost tracking of objects at $t$ and $t+1$, Acceleration $V_{12} - V_{01}$ is used.}
\label{fig:1}       
\end{figure}

\subsection*{3. Simulation Analyses}
\label{subsec:4}
To test our algorithm, we conducted three simulation analyses and compared speed cost with acceleration cost. As mentioned, a larger value of $\lambda$ leads to a wider search range. Therefore, calculation may become difficult if $\lambda$ is very large. We thus set appropriate values (10 or 100) for each simulation. For the calculation using the original Sinkhorn method, we used the Python Optimal Transport Library package \cite{pot}.

Initially, we generated simulation data with multiple objects that moved at constant velocity without noise (Simulation 1). The initial coordinate values of x and y were sampled from a normal distribution with a mean of 0 and a variance of 1. Speed along the x and y coordinates was determined by multiplying a random number sampled from this normal distribution by parameter $m$. Speeds of less than 0 were set to 0 to create some objects that were at rest along the x or y coordinates. Here, $m$ is a parameter for adjusting speed. First, we compared the performance of tracking based on speed cost and acceleration cost, where $m=0.5$ and $n=100$. Then, we changed the number of objects $n$ (50 or 200) and parameter $m$ (0.5 or 2.0) to evaluate their influence.

Next, we investigated the performance obtained using a third cost array in tracking objects moving at constant velocity (Simulation 2). 
We conducted two types of tracking, namely Acceleration (2D) and Acceleration (3D).
The procedure of Acceleration (2D) was as follows. First, the correspondence between $t$ and $t+1$ was determined using speed cost tracking. Then, assuming constant velocity, the position at $t+2$ was predicted, and the nearest object was associated. For Acceleration (3D), the third cost array was used, as done for Simulation 1. However, we compressed the information in the triples down to $p'_{ik} = \sum_{j=1}^n p_{ijk}$ instead of $p'_{ij} = \sum_{k=1}^n p_{ijk}$. The accuracy of tracking at $t+2$ was compared between Acceleration (2D) and Acceleration (3D). The parameter settings were $m=2.0$ and $n=100$.

Next, we considered random-walking objects (Simulation 3). The movement distance at each time point was sampled from the 2D normal distribution with mean vector $\bf0$ and variance matrix $\sigma^2\bf{I}$. Values of $\sigma^2$ were varied (0.1, 0.5, 1.0, 1.5, 2.0) and $n$ was set to 100.

Finally, we considered multiple objects that move at constant velocity with noise (Simulation 4). The settings were the same as those for Simulation 1, where x and y coordinate values were sampled from the normal distribution with a mean of 0 and a variance of 1. Speed along the x and y coordinates was determined by multiplying a random number sampled from this normal distribution by parameter $m$. Parameters $n$ and $m$ were set to 100 and 0.5, respectively. Unlike in Simulation 1, in Simulation 3, noise was added to both x and y coordinate values each time there was movement. The noise at each time point was sampled from the 2D normal distribution with mean vector $\bf0$ and variance matrix $\sigma^2\bf{I}$. We used four values of $\sigma^2$ (0.01, 0.05, 0.10, 0.25).

We evaluated the tracking performance using a performance index whose values range between 0 and 1. Because a given object has the same index in our simulations, the correspondence of correct answers lies diagonally in the optimal transport matrix. Thus, the performance index is defined as the number of rows whose diagonal value is maximum. A higher index value represents better tracking performance. In all simulations, 10 datasets were generated, and the performance index was calculated for each dataset.

\section{Results}
Figure 2 shows the performance index results for speed cost and acceleration cost obtained with $n=100$ and $m=0.5$ (Simulation 1). The performance index for acceleration cost was better than that for speed cost. Next, the same simulation was performed by changing $n$. Figure 3 shows the performance index values obtained with $n=50$ and $n=200$. For both speed cost and acceleration cost, as $n$ increased, the performance index decreased. Figure 4 shows the results of the same simulation with $m$ changed ($m=0.5$ and $m=2.0$) and $n$ set to 50. For speed cost, but not for acceleration cost, the performance index was affected by $m$. \ref{tab:2} shows the average performance index values for each parameter setting in Simulation 1. Figure 5 shows the performance index results for Acceleration (2D) and Acceleration (3D) obtained with $n=100$ and $m=2.0$ (Simulation 2). Although constant-velocity motion was assumed for both cases, performance was better when using the 3D cost array (Acceleration (3D)).

\begin{figure}[tbp]
\includegraphics[scale=.7]{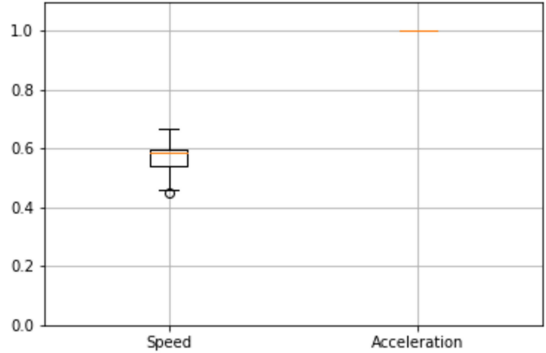}
%
%
\captionsetup{labelformat=empty,labelsep=none}
\caption{Fig. 2: Boxplot of performance index for 10 simulation datasets for speed cost and acceleration cost tracking of objects moving at constant velocity without noise (Simulation 1). Parameters $n$ and $m$ were set to 100 and 0.5, respectively. The boxes show the lower quantile and upper quartile values of the data. The orange lines represent the median values. The whiskers and white circle are the range of data (minimum value to maximum value) and outlier value, respectively, which were defined by the default settings of the matplotlib.pyplot.boxplot function in the Python library.}
\label{fig:1}       
\end{figure}

\begin{figure}[tbp]
\includegraphics[scale=.5]{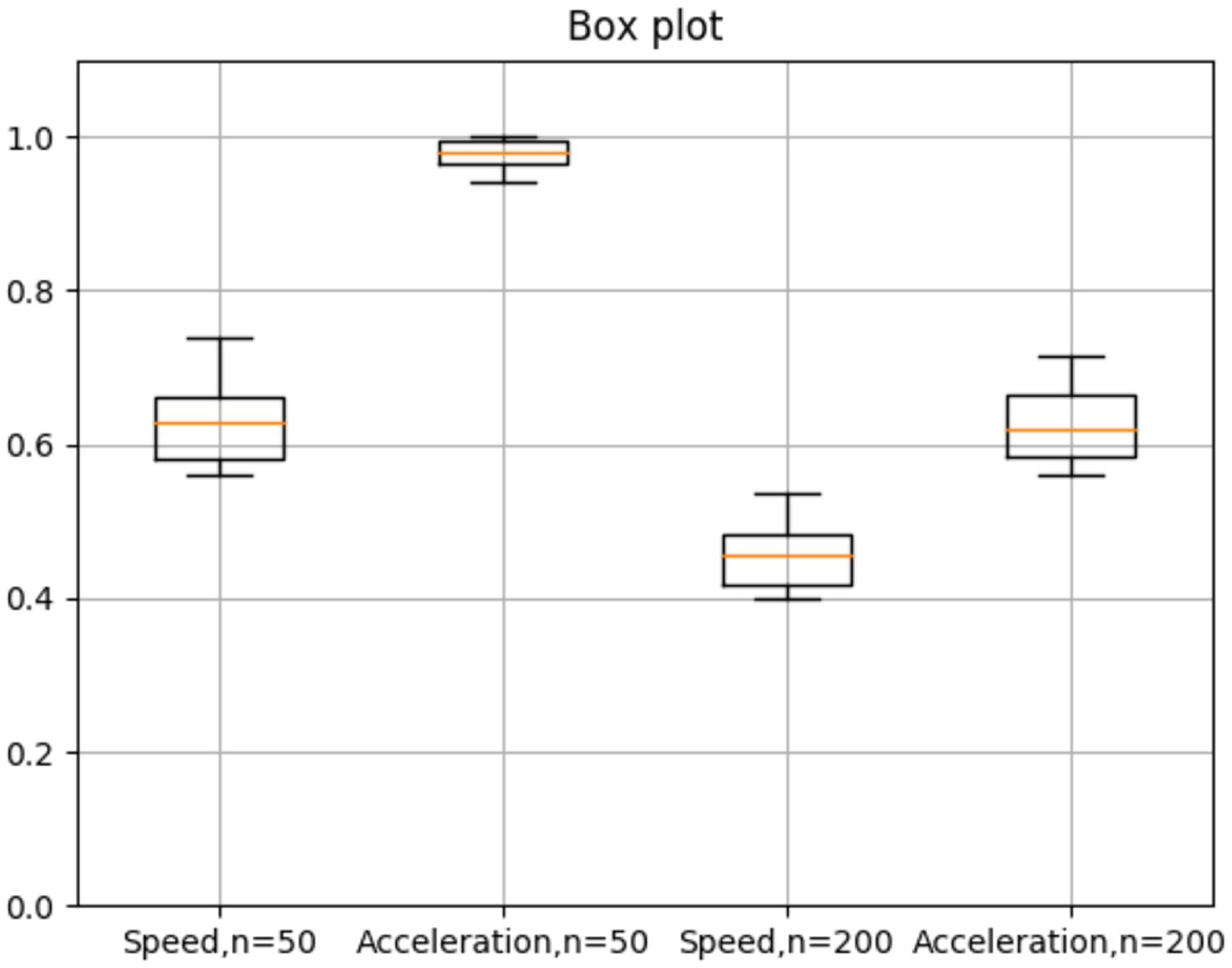}
\centering
%
\captionsetup{labelformat=empty,labelsep=none}
\caption{Fig. 3: Boxplot of performance index for 10 simulation datasets for speed cost and acceleration cost tracking of objects moving at constant velocity without noise (Simulation 1) obtained for $n=50$ and $n=200$. Parameter $m$ was set to 0.5. The description of the objects in the boxplot is the same as Figure 2.}
\label{fig:2}       
\end{figure}

\begin{figure}[tbp]
\includegraphics[scale=.45]{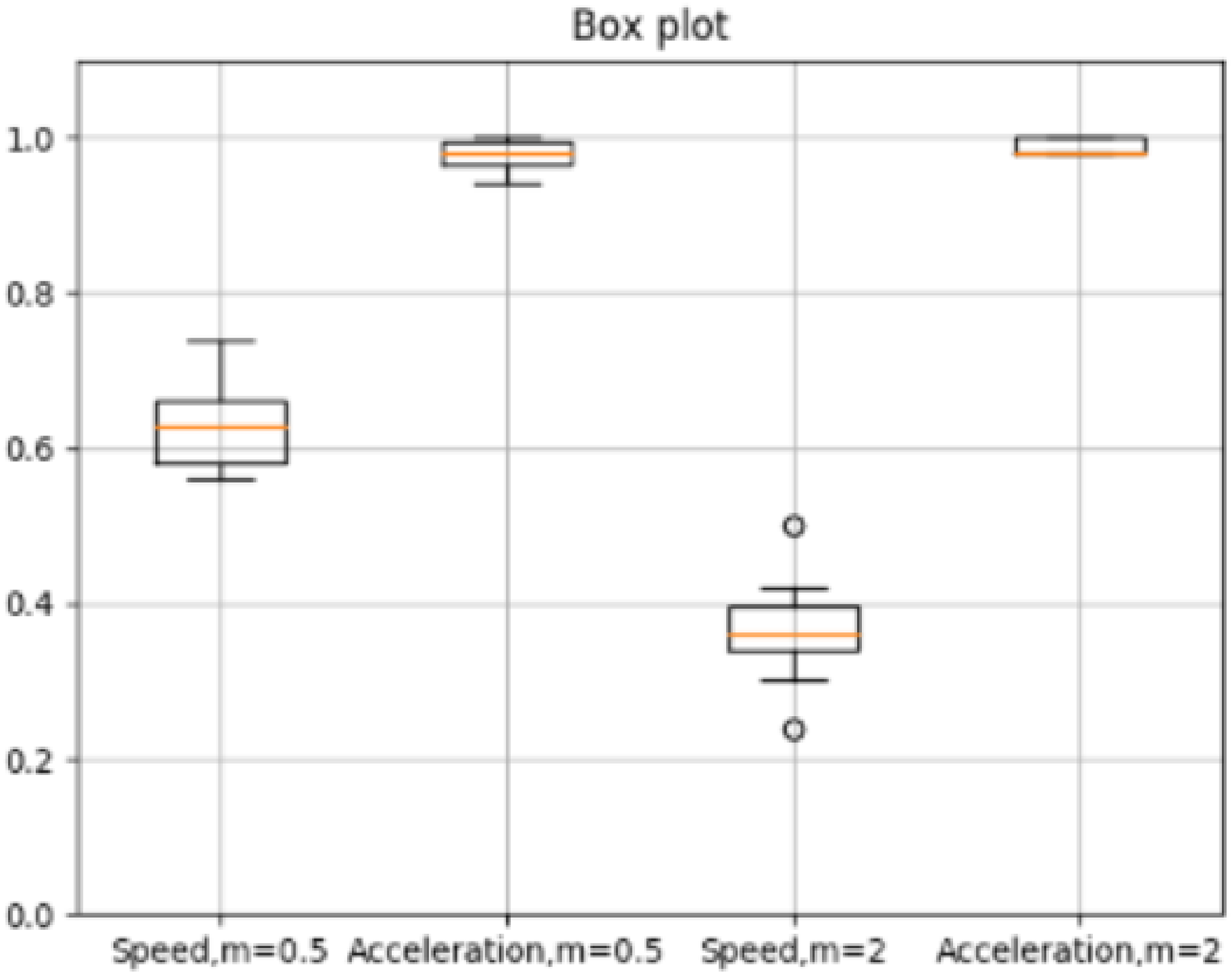}
\centering
%
\captionsetup{labelformat=empty,labelsep=none}
\caption{Fig. 4: Boxplot of performance index for 10 simulation datasets for speed cost and acceleration cost tracking of objects moving at constant velocity without noise (Simulation 1) obtained with $m=0.5$ and $m=2.0$. Parameter $n$ was set to 50. The description of the objects in the boxplot is the same as Figure 2.}
\end{figure}

\begin{figure}[tbp]
\includegraphics[scale=.5]{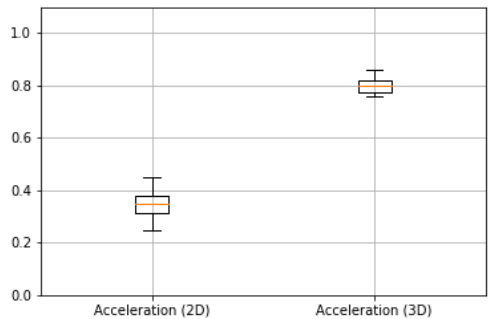}
\centering
%
\captionsetup{labelformat=empty,labelsep=none}
\caption{Fig. 5: Boxplot of performance index for 10 simulation datasets for tracking with acceleration tracking based on 2D Sinkhorn (Acceleration (2D)) and 3D Sinkhorn (Acceleration (3D)) (Simulation 2). The description of the objects in the boxplot is the same as Figure 2.}
\label{fig:4}       
\end{figure}

%
\begin{table}
\caption{Average performance index for each parameter setting in Simulation 1}
\label{tab:2}       
%
%
\begin{tabular}{p{1cm}p{1cm}p{2cm}p{2cm}}
\hline\noalign{\smallskip}
$n$ & $m$ & Speed cost & Acceleration cost  \\
\hline\noalign{\smallskip}
100 &  0.5 & 0.567 & 1.0  \\
50 & 0.5 & 0.628 &  0.976 \\
200 & 0.5 &  0.456 &  0.624  \\
50 & 2 & 0.364 & 0.988  \\
\hline\noalign{\smallskip}
\end{tabular}
\end{table}

Figure 6 shows the results of Simulation 3. The performance index for tracking random-walking objects was generally low for both speed cost and acceleration cost. Figure 7 shows the results of Simulation 4. The performance index decreased as noise increased for both speed cost and acceleration cost. The results suggest that if the added noise is not excessively large, acceleration cost tracking outperforms speed cost tracking. However, when the added noise is large, speed cost tracking is slightly better. 

%
\begin{figure}[tbp]
\includegraphics[scale=.7]{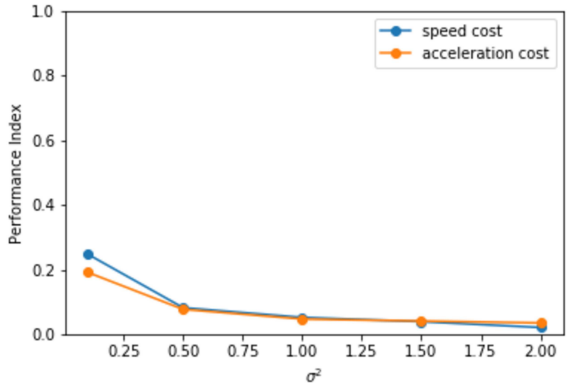}
%
%
\captionsetup{labelformat=empty,labelsep=none}
\caption{Fig. 6: Accuracy of tracking random-walking objects (Simulation 3). The movement distance at each time point was sampled from the 2D normal distribution with mean vector $\bf0$ and variance matrix $\sigma^2\bf{I}$. Values of $\sigma^2$ were changed (0.1, 0.5, 1.0, 1.5, 2.0). Parameter $n$ was set to 100. The plots show the average performance index for 10 datasets for speed cost and 10 datasets for acceleration cost.}
\label{fig:5}       
\end{figure}

%
\begin{figure}[tbp]
\includegraphics[scale=.7]{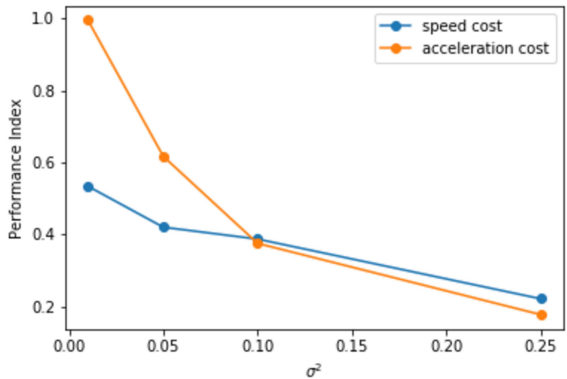}
\captionsetup{labelformat=empty,labelsep=none}
\caption{Fig. 7: Accuracy of tracking objects moving at constant velocity with noise (Simulation 4). Here, $\sigma^2$ is the variance of noise. The settings are the same as those for Simulation 1, where x and y coordinates were sampled from the normal distribution with a mean of 0 and a variance of 1. Speed along the x and y coordinates was determined by multiplying a random number sampled from the normal distribution by parameter $m$. Parameters $n$ and $m$ were set to 100 and 0.5, respectively. The plots show the average performance index for 10 datasets for speed cost and 10 datasets for acceleration cost.
}
\label{fig:6}       
\end{figure}

\section{Discussion}
For tracking an agent that is moving at constant velocity, the results of Simulation 1 show that the performance obtained using acceleration cost tracking is better and that it is not affected by speed. Although performance deteriorates as the number of objects increases for both tracking methods, acceleration cost tracking still shows better performance. The results of Simulation 2 show that the Sinkhorn method with a 3D cost array is better than that with a second cost matrix for tracking objects moving at constant velocity. The results of Simulation 3 show that neither speed cost tracking nor acceleration cost tracking are useful for random-walking objects. They also show that when objects move at constant velocity, our method is superior to the nearest-neighbor method, even if small noise is added. However, when the added noise is large, speed cost tracking is slightly better. Thus, the above results indicate that for objects moving at constant velocity, particularly when the movement is intense, the number of objects is large, and the added noise is not excessively large, the proposed method of acceleration cost tracking is superior to speed cost tracking based on the conventional nearest-neighbor method.

The advantage of using the Sinkhorn distance for tracking is that various assumptions regarding movement can be incorporated into cost matrix $M$. In this study, we proposed a tracking method based on the assumption of uniform object motion. By considering the third difference, the difference in acceleration, we can also express the assumption of equal acceleration motion. In addition, if appropriate cost arrays can be set, it may be possible to apply the proposed method to objects that move smoothly, such as amoeba cells. Therefore, the proposed method is potentially useful for bioimaging research.

Because of the flexibility of the cost matrix and cost array, the Sinkhorn method can consider features other than motion characteristics. For example, each object in an image has a specific shape. Changes in shape can be used for the cost. Moreover, multiple costs, such as the costs of shape, distance, and acceleration, can be combined to define an optimization cost. Our study considered just one of many potential applications of the Sinkhorn method to the tracking problem.

\section{Conclusion}
We proposed a method that applies the Sinkhorn distance to track objects. We compared speed cost tracking based on the conventional nearest-neighbor method and acceleration cost tracking based on the proposed Sinkhorn method, and compared their performance using simulation data. We showed that the Sinkhorn method can be applied effectively to object tracking. Our simulation data analysis suggests that when objects are moving at constant velocity, our method, which sets acceleration as a cost, outperforms the traditional nearest-neighbor method in terms of tracking objects as long as the added noise is not excessively large. To apply the proposed method to real bioimaging data, it is necessary to set an appropriate cost indicator based on the movement features.

\section{Acknowledgement}
This work was supported by Grant-in-Aid for JSPS KAKENHI Grant Number JP19J14816, and Core Research for Evolutional Science and Technology (CREST) (grant numbers JPMJCR1502 and JPMJCR15G1) of the Japan Science and Technology Agency (JST). We would like to thank to the members of Matsuda Laboratory, Osaka University Graduate School of Information Science, and Ishi Laboratory, Osaka University Graduate School of Medicine, and Dr. Kazushi Mimura of Hiroshima City University for useful advice and discussion. The authors declare no conflicts of interest.

\bibliography{mybibfile} 
\bibliographystyle{junsrt} 

\end{document}